\documentclass[universe,article,accept,pdftex,moreauthors]{Definitions/mdpi} 
\graphicspath{{Definitions/Fig/}}

\firstpage{1} 
\pubvolume{9}
\issuenum{11}
\articlenumber{467}
\pubyear{2023}
\copyrightyear{2023}
\externaleditor{Academic Editors: Ana Alonso-\linebreak Serrano, Matt Visser, Jessica Santiago and Sebastian Schuster} 
\datereceived{14 September 2023} 
\daterevised{24 October 2023} 
\dateaccepted{26 October 2023} 
\datepublished{30 October 2023} 
\hreflink{https://doi.org/\linebreak 10.3390/universe9110467} 

\Title{Closed Timelike Curves Induced by a Buchdahl-Inspired Vacuum Spacetime in $\mathcal{R}^{2}$ Gravity}

\TitleCitation{Closed Timelike Curves Induced by a Buchdahl-Inspired Vacuum Spacetime in $\mathcal{R}^{2}$ Gravity}

\Author{Hoang Ky Nguyen $^{1,}$*\orcidA{} and Francisco S. N. Lobo $^{2,3}$\orcidB{}}

\AuthorNames{Hoang Ky Nguyen and Francisco S. N. Lobo}

\AuthorCitation{Nguyen, H.K.; Lobo, F.S.N.}

\address{%
$^{1}$ \quad  Department of Physics, Babe\c{s}–Bolyai University, 400084 Cluj-Napoca, Romania\\
$^{2}$ \quad Instituto de Astrof\'{\i}sica e Ci\^{e}ncias do Espa\c{c}o, Faculdade de
Ci\^encias da Universidade de Lisboa, \mbox{Campo Grande}, Edif\'{\i}cio C8,
1749-016 Lisbon, Portugal; fslobo@fc.ul.pt\\
$^{3}$ \quad Departamento de F\'{i}sica, Faculdade de Ci\^{e}ncias da Universidade de Lisboa, Campo Grande, Edif\'{\i}cio C8, 1749-016 Lisbon, Portugal}

\corres{Correspondence: hoang.nguyen@ubbcluj.ro}

\abstract{The recently obtained \emph{special} Buchdahl-inspired metric [Phys. Rev. D 107, 104008 (2023)] describes asymptotically flat spacetimes in pure Ricci-squared gravity.
The metric depends on a new (Buchdahl) parameter $\tilde{k}$ of higher-derivative
characteristic, and reduces to the Schwarzschild metric, for $\tilde{k}=0$.
For the case $\tilde{k}\in(-1,0)$, it was shown that it describes
a traversable Morris--Thorne--Buchdahl (MTB) wormhole [Eur. Phys. J. C 83, 626 (2023)], where the weak energy condition is formally violated. In this paper, we briefly review the \emph{special} Buchdahl-inspired metric, with focuses on the construction of the Kruskal--Szekeres (KS) diagram and the
situation for a wormhole to emerge. Interestingly, the MTB wormhole
structure appears to permit the formation of closed timelike curves
(CTCs). More specifically, a CTC straddles the throat, comprising
of two segments positioned in opposite quadrants of the KS
diagram. The closed timelike loop thus passes through the wormhole
throat twice, causing \emph{two} reversals in the time direction experienced by the (timelike) traveller on the CTC. The key to constructing a
CTC lies in identifying any given pair of antipodal points $(T,X)$
and $(-T,-X)$ \emph{on the wormhole throat} in the KS diagram
as corresponding to the same spacetime event. It is interesting to
note that the Campanelli--Lousto metric in Brans--Dicke gravity is known
to support two-way traversable wormholes, and the formation of the
CTCs presented herein is equally applicable to the Campanelli--Lousto
solution.
}


\begin{document}


\section{Introduction}

Wormholes are hypothetical shortcuts in spacetime, and~are solutions
for the gravitational field equations, where the fundamental ingredient
is the flaring-out condition~\citep{MorrisThorne-1988-1}. In~classical
General Relativity, the~latter condition entails the violation of
the null energy condition, and~consequently all of the energy conditions
\citep{Visser-book,Alcubierre:2017pqm}. However, it has been shown
that in modified theories of gravity, the matter threading the wormhole
throat may satisfy the energy conditions, and~it is the higher order
curvature terms, interpreted as a gravitational fluid, that support
these nonstandard wormhole geometries~\citep{Lobo:2009ip,Harko:2013yb}.
Another extremely interesting feature of traversable wormholes is
that they may be hypothetically manipulated to induce closed timelike
curves (CTCs)~\citep{MorrisThorne-1988-2,Novikov-CTCWH,frolovnovikovTM,CauchyCTC}.
In fact, General Relativity is contaminated with non-trivial geometries,
which generate CTCs~\citep{Godel,Pfarr,Tipler-CTCs,Felice,GottCTC,Ori,Ori:2005ht,Alcubierre,EverettCTC},
which allow time travel, in~the sense that an observer that travels
on a trajectory in spacetime along this curve and may return to an event
before his departure~\citep{Lobo-2008}.

Due to the interesting physics involved in these exotic spacetimes,
much attention has been given to these geometries in the literature,
and we refer the reader to~\citep{Alcubierre:2017pqm}, and~the references
therein, for~a recent review. As~wormhole physics has been explored
extensively in modified theories of gravity, where the higher-order curvature terms support these wormhole geometries~\citep{Harko:2013yb},
in this work, we shall be interested in the recent \emph{special}
Buchdahl-inspired metric, obtained by one of the present authors~\citep{Nguyen-2022-Lambda0},
that describes asymptotically flat spacetimes in pure $\mathcal{R}^{2}$
gravity~\citep{Buchdahl-1962}.~The~metric is dependent on a new (Buchdahl)
parameter $\tilde{k}$ of higher-derivative characteristic, and~recovers
the Schwarzschild metric when $\tilde{k}=0$. In~a recent work~\citep{2023-WH},
it was demonstrated that the \emph{special} Buchdahl-inspired metric
supports a two-way traversable Morris--Thorne--Buchdahl (MTB) wormhole
for $\tilde{k}\in(-1,0)$, in~which case the weak energy condition
is formally~violated.

In this paper, we shall review the \emph{special} Buchdahl-inspired
metric, with~focuses on the construction of the $\zeta$--Kruskal--Szekeres
(KS) diagram and the conditions to generate a wormhole. Curiously,
the MTB wormhole structure appears to permit the formation of CTCs. 
A CTC straddles the throat, comprising of
two segments positioned in the opposite Quadrant I and Quadrant III
of the $\zeta$--KS diagram. The~closed timelike loop thus passes through
the wormhole throat twice, causing \emph{two} reversals in the time
direction experienced by the (timelike) traveller on the CTC. The~
key to constructing a CTC lies in identifying any given pair of antipodal
points $(T,X)$ and $(-T,-X)$ \emph{on the wormhole throat} in the
$\zeta$--KS diagram as corresponding to the same spacetime~event.

In a previous work~\citep{Poplawski-2010}, Pop\l{}awski put forth
this procedure for the Einstein--Rosen bridge in which he identified
antipodal points \emph{on the horizon} as corresponding to the same
spacetime event. Although~his maneuver may prove untenable for ``Schwarzschild
wormholes'' (and he did not report a CTC), we adapt his construction
to the MTB wormhole, with~a minor but crucial \emph{modification}:
instead of the horizon, the~identification of antipodal points takes
place on the wormhole throat which permits a two-way traversal. In~
certain~situations, the~Campanelli--Lousto solution in Brans--Dicke
gravity is known to support two-way traversable wormholes~\citep{2023-WEC}.
When this occurs, the~formation of CTCs presented herein is equally
applicable to the Campanelli--Lousto solution. Generally speaking,
our CTC should be a generic aspect of the family of scalar--tensor
theories.

This paper is organized in the following manner: In Section~\ref{sec:overview}, we present the vacuum \emph{special} Buchdahl-inspired metric, and~in Sections~\ref{sec:KS-coord} and  \ref{sec:KS-diagram}, we review
the causal structure of the solution, using the Kruskal--Szekeres diagram,
which is a maximal analytic extension of the \emph{special} Buchdahl-inspired
metric. For~completeness, Section~\ref{sec:naked} provides a brief
detour for the case of naked singularities. Section~\ref{sec:CTC},
the most essential one, describes our construction of the CTCs for
the MTB wormholes. Finally, in~Section~\ref{sec:discussions}, we discuss
our results and~conclude.

\section{The \emph{Special} Buchdahl-Inspired Metric: Brief~Review}

\label{sec:overview}

The pure $\mathcal{R}^{2}$ action, $\int d^{4}x\sqrt{-g}\,\mathcal{R}^{2}$,
yields the field equation in vacuo~\citep{Buchdahl-1962}
\begin{equation}
\mathcal{R}\left(\mathcal{R}_{\mu\nu}-\frac{1}{4}g_{\mu\nu}\mathcal{R}\right)+g_{\mu\nu}\square\mathcal{R}-\nabla_{\mu}\nabla_{\nu}\mathcal{R}=0\,.
\end{equation}
Despite the fourth-order nature of this equation, in~\citep{Nguyen-2022-Lambda0},
one of the present authors obtained an exact closed analytical solution,
named the \emph{special} Buchdahl-inspired metric. It describes a
static and spherically symmetric vacuum configuration
\begin{equation}
ds^{2}=\left|1-\frac{r_{\text{s}}}{r}\right|^{\tilde{k}}\left\{ -\left(1-\frac{r_{\text{s}}}{r}\right)dt^{2}+\left(\frac{\rho(r)}{r}\right)^{4}\frac{\,dr^{2}}{1-\frac{r_{\text{s}}}{r}}+\left(\frac{\rho(r)}{r}\right)^{2}\,r^{2}d\Omega^{2}\right\} \,,\label{eq:B-metric-1}
\end{equation}
where the function $\rho(r)$ is given by virtue of
\begin{equation}
\left(\frac{\rho(r)}{r}\right)^{2}:=\frac{\zeta^{2}\left|1-\frac{r_{\text{s}}}{r}\right|^{\zeta-1}}{\left(1-\text{s}\left|1-\frac{r_{\text{s}}}{r}\right|^{\zeta}\right)^{2}}\,\left(\frac{r_{\text{s}}}{r}\right)^{2}\,.\label{eq:B-metric-2}
\end{equation}
The dimensionless parameters $\tilde{k}$ and $\zeta$ are defined
as $\tilde{k}:=k/r_{\text{s}}$, $\zeta:=\sqrt{1+3\tilde{k}^{2}}$,
and $\text{s}:=\pm1$ denotes the {\it signum} of $1-\frac{r_{\text{s}}}{r}$.
Here, $\tilde{k}$ is a new (Buchdahl) parameter of higher-derivative
characteristic, and~$r_{\text{s}}$ plays the role of a Schwarzschild
radius. At~$\tilde{k}=0$, \mbox{$\rho(r)\equiv r$} and \mbox{Equation~\eqref{eq:B-metric-1}}
recovers the Schwarzschild metric. At~spatial infinity, metric \eqref{eq:B-metric-1}
is asymptotically flat (Note: a more general solution expressed in
a compact form, named the Buchdahl-inspired metric, was obtained by
one of the authors in Refs.~\citep{Nguyen-2022-Buchdahl,Nguyen-2023-essay}
by completing the original but unfinished work of Buchdahl~\citep{Buchdahl-1962}.
This latter solution is asymptotically de Sitter, and~is specified
by \emph{four} parameters, reflecting the fourth-derivative nature
of a quadratic theory).

Upon another coordinate transformation~\citep{2023-WH}\vspace{-6pt}
\begin{equation}
1-\frac{r'_{\text{s}}}{r'}=\text{s}\left|1-\frac{r_{\text{s}}}{r}\right|^{\zeta}\,,
\end{equation}
in which $r'_{\text{s}}:=\zeta\,r_{\text{s}}$ and $\text{\text{s}}=\text{sgn}\left(1-\frac{r_{\text{s}}}{r}\right)=\text{sgn}\left(1-\frac{r'_{\text{s}}}{r'}\right)$,
the metric given in Equations~\eqref{eq:B-metric-1} and \eqref{eq:B-metric-2}
can be brought into the following form\vspace{-6pt}
\begin{equation}
ds^{2}=-\text{s}\left|1-\frac{r'_{\text{s}}}{r'}\right|^{A}dt^{2}+\text{s}\left|1-\frac{r'_{\text{s}}}{r'}\right|^{B}dr'^{2}+\left|1-\frac{r'_{\text{s}}}{r'}\right|^{B+1}r'^{2}d\Omega^{2}\,,\label{eq:CL}
\end{equation}
where
\begin{equation}
A:=\frac{\tilde{k}+1}{\zeta},\qquad B:=\frac{\tilde{k}-1}{\zeta},\qquad\zeta:=\sqrt{1+3\tilde{k}^{2}}\,.\label{eq:A-B-special}
\end{equation}
The parameters satisfy the following relation:\vspace{-6pt}
\begin{equation}
A^{2}+AB+B^{2}=1\,.\label{eq:A-B-relation}
\end{equation}

It is worth noting that the metric expressed in Equation~\eqref{eq:CL}
also describes the \emph{generalized} Campanelli--Lousto (CL) solution
for the Brans--Dicke action, $\int d^{4}x\sqrt{-g}\left[\phi\,\mathcal{R}-\frac{\omega}{\phi}\nabla^{\mu}\phi\nabla_{\mu}\phi\right]$.
In the \emph{generalized} CL metric, which was found recently in~\citep{2023-WEC}
by one of the present authors, $A$ and $B$ take on any value in
$\mathbb{R}$, and~are linked by\vspace{-6pt}
\begin{equation}
A^{2}+AB+B^{2}-1=-\frac{\omega}{2}(A+B)^{2}\,,
\end{equation}
with $\omega$ being the Brans--Dicke~parameter.

In the rest of this work, we shall concern ourselves with the \emph{special}
Buchdahl-inspired metric in pure $\mathcal{R}^{2}$ gravity. Dropping
the prime in the notation of $r'$ and $r'_{\text{s}}$ in Equation~\eqref{eq:CL},
we shall use the explicit expression below\vspace{-6pt}
\begin{equation}
ds^{2}=\left|1-\frac{r_{\text{s}}}{r}\right|^{\frac{\tilde{k}}{\zeta}}\left\{ -\text{s}\left|1-\frac{r_{\text{s}}}{r}\right|^{\frac{1}{\zeta}}dt^{2}+\text{s}\left|1-\frac{r_{\text{s}}}{r}\right|^{-\frac{1}{\zeta}}dr^{2}+\left|1-\frac{r_{\text{s}}}{r}\right|^{1-\frac{1}{\zeta}}r^{2}d\Omega^{2}\right\} \,,\label{eq:B-metric-new}
\end{equation}
where
\begin{equation}
\text{s}:=\text{sgn}\left(1-\frac{r_{\text{s}}}{r}\right),\qquad\zeta:=\sqrt{1+3\tilde{k}^{2}}\,.
\end{equation}

\section{The \boldmath{$\zeta$}--Kruskal--Szekeres~Coordinates}

\label{sec:KS-coord}

The construction of the KS diagram for the \emph{special} Buchdahl-inspired
metric has been carried out in Refs.~\citep{Nguyen-2022-Lambda0,2023-WEC}.
However, for~self-consistency and self-completeness, we summarize
the key points~here.
\begin{itemize}
\item The tortoise coordinate $r^{*}(r)$ for the \emph{special} Buchdahl-inspired
metric \eqref{eq:B-metric-new} is defined by virtue of
\begin{equation}
dr^{*}=\frac{\text{s}}{\left|1-\frac{r_{\text{s}}}{r}\right|^{1/\zeta}}\,dr\,.\label{eq:tortois-eqn}
\end{equation}
This equation is soluble, yielding the tortoise coordinate in terms
of a Gaussian hypergeometric function\vspace{-6pt}

\begin{align}
r^{*}(r) & =\frac{r_{\text{s}}}{1-1/\zeta}\,\left|1-\frac{r_{\text{s}}}{r}\right|^{1-1/\zeta}\,_{2}F_{1}\left(2,1-1/\zeta;2-1/\zeta;1-\frac{r_{\text{s}}}{r}\right)-\frac{r_{\text{s}}\,\pi/\zeta}{\sin(\pi/\zeta)}\,,\label{eq:tortoise}
\end{align}
which is represented in Figure~\ref{fig:Tortoise-coordinate}.
The integration constant required for Equation~\eqref{eq:tortois-eqn}
has been chosen such that $r^{*}=0$ when $r=0$.

\begin{figure}[H]
\includegraphics[scale=0.7]{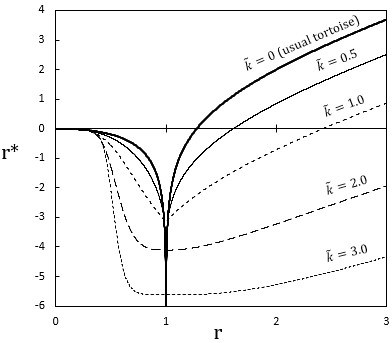} \caption{\label{fig:Tortoise-coordinate}The $\zeta$--tortoise coordinate given
by Equation~\eqref{eq:tortoise} for various values of $\tilde{k}$ (with
$r_{\text{s}}=1$).}
\end{figure}

\item The advanced and retarded Eddington--Finkelstein coordinates are defined
as
\begin{align}
v & :=t+r^{*}\,,\\
u & :=t-r^{*}\,,
\end{align}
respectively.\vskip6pt
\item For the Kruskal--Szekeres (KS) coordinates, it is necessary to separate
the two ranges, $r>r_{\text{s}}$ versus $r<r_{\text{s}}$.
\begin{itemize}
\item For $r>r_{\text{s}}$, we define
\begin{align}
X & :=\frac{1}{2}\left(e^{\frac{v}{2r_{\text{s}}}}+e^{-\frac{u}{2r_{\text{s}}}}\right)=e^{\frac{r^{*}(r)}{2r_{\text{s}}}}\cosh\frac{t}{2r_{\text{s}}}\,,\label{eq:X-ex}\\
T & :=\frac{1}{2}\left(e^{\frac{v}{2r_{\text{s}}}}-e^{-\frac{u}{2r_{\text{s}}}}\right)=e^{\frac{r^{*}(r)}{2r_{\text{s}}}}\sinh\frac{t}{2r_{\text{s}}}\,.\label{eq:T-ex}
\end{align}
\item For $r<r_{\text{s}}$, we define
\begin{align}
X & :=\frac{1}{2}\left(e^{\frac{v}{2r_{\text{s}}}}-e^{-\frac{u}{2r_{\text{s}}}}\right)=e^{\frac{r^{*}(r)}{2r_{\text{s}}}}\sinh\frac{t}{2r_{\text{s}}}\,,\label{eq:X-in}\\
T & :=\frac{1}{2}\left(e^{\frac{v}{2r_{\text{s}}}}+e^{-\frac{u}{2r_{\text{s}}}}\right)=e^{\frac{r^{*}(r)}{2r_{\text{s}}}}\cosh\frac{t}{2r_{\text{s}}}\,.\label{eq:T-in}
\end{align}
\end{itemize}
\end{itemize}
In combination, the~\emph{special} Buchdahl-inspired metric in the
Kruskal--Szekeres (KS) coordinates is thus
\begin{equation}
ds^{2}=\left|1-\frac{r_{\text{s}}}{r}\right|^{\frac{\tilde{k}}{\zeta}}\biggl\{-4r_{\text{s}}^{2}e^{-\frac{r^{*}}{r_{\text{s}}}}\left|1-\frac{r_{\text{s}}}{r}\right|^{\frac{1}{\zeta}}\left(dT^{2}-dX^{2}\right)+r^{2}\left|1-\frac{r_{\text{s}}}{r}\right|^{1-\frac{1}{\zeta}}d\Omega^{2}\biggr\}\,,\label{eq:KS-metric}
\end{equation}
where\vspace{-6pt}
\begin{align}
T^{2}-X^{2} & =-\text{s}\;e^{\frac{r^{*}(r)}{r_{\text{s}}}}\nonumber \\
 & =-\text{s}\;\exp\left[\frac{\left|1-\frac{r_{\text{s}}}{r}\right|^{1-1/\zeta}}{1-1/\zeta}\,_{2}F_{1}\left(2,1-1/\zeta;2-1/\zeta;1-\frac{r_{\text{s}}}{r}\right)-\frac{\pi/\zeta}{\sin(\pi/\zeta)}\right]\,,\label{eq:T-X-1}\\
\frac{T}{X} & =\left(\tanh\frac{t}{2r_{\text{s}}}\right)^{\text{s}}\,,\label{eq:T-X-2}
\end{align}
with $\text{s}=\pm1$ for the ranges $r\in(r_{\text{s}},\infty)$
and $r\in(-\infty,r_{\text{s}})$, respectively. Note that in this
final form, the~coordinates $T$ and $X$ are \emph{enlarged} to satisfy
Equations~\eqref{eq:T-X-1} and \eqref{eq:T-X-2}. That is to say, each
pair of coordinates $(t,r)$ corresponds to \emph{two} pairs of coordinates
$(T,X)$ and $(-T,-X)$, with~$X$ and $T$ given by Equations~\eqref{eq:X-ex} and \eqref{eq:T-ex}
for the range $r\in(r_{\text{s}},\infty)$ and Equations~\eqref{eq:X-in} and \eqref{eq:T-in}
for the range $r\in(-\infty,r_{\text{s}})$.

It is apt and convenient to call the range $r\in(r_{\text{s}},\infty)$
an ``exterior'' region. However, since the Kretschmann invariant in
general diverges at $r=r_{\text{s}}$~\citep{Nguyen-2022-Lambda0},
we shall avoid naming the range $r\in(0,r_{\text{s}})$ an ``interior''
region, from~here~on.

\section{The \boldmath{$\zeta$}-Kruskal--Szekeres~Diagram}

\label{sec:KS-diagram}

Restricting within the $(T,X)$ plane, i.e.,~$d\theta=d\varphi=0$,
the $\zeta$--KS diagram for metric~\eqref{eq:KS-metric} is shown
in Figure~\ref{fig:KS-diagram}. We refer to a number of key features,
developed in Ref.~\citep{Nguyen-2022-Lambda0}:
\vspace{-6pt}

\begin{figure}[H]
\includegraphics[scale=0.68]{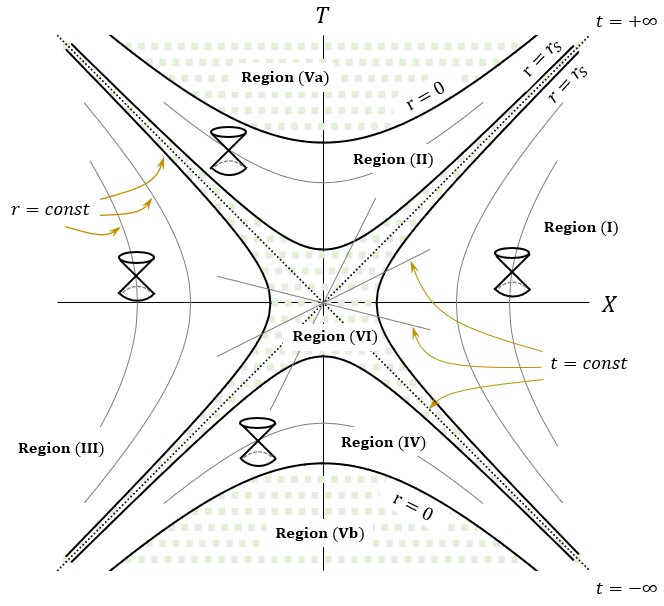}\vspace{-6pt}
\caption{\label{fig:KS-diagram}$\zeta$--Kruskal--Szekeres diagrams for the
\emph{special} Buchdahl-inspired spacetimes, given by \mbox{Equation~\eqref{eq:B-metric-1}}
or Equation~\eqref{eq:B-metric-new}. Each point on the diagram is a 2-sphere. {Except for $\tilde{k}=0$ and $\tilde{k}=-1$, the~Kretschmann scalar
diverges on the hyperbolae $r=r_{\text{s}}$ and $r=0$; see Appendix \ref{sec:Kretschmann-scalar} for~exposition.}}
\end{figure}

\begin{enumerate}
\item The $\zeta$--KS diagram is conformally Minkowski. The~null geodesics
are $dX=\pm dT$. 
\item Per Equations~\eqref{eq:T-X-1} and \eqref{eq:T-X-2}, a~constant--$r$
contour corresponds to a hyperbola, while a constant--$t$ contour
corresponds to a straight line running through the origin of the $(T,X)$
plane. The~coordinate origin $r=0$ amounts to $T^{2}-X^{2}=1$, since
$r^{*}(r=0)=0$. 
\item The boundary $r=r_{\text{s}}$ corresponds to \emph{two} distinct
hyperbolae, given by\vspace{-6pt}
\begin{equation}
T^{2}-X^{2}=\begin{cases}
-e^{-\frac{\pi/\zeta}{\sin(\pi/\zeta)}} & \text{for }r>r_{\text{s}}\,,\\
+e^{-\frac{\pi/\zeta}{\sin(\pi/\zeta)}} & \text{for }r<r_{\text{s}}\,.
\end{cases}\label{eq:hyperbolae}
\end{equation}
Since each hyperbola has two separate branches on its own, Figure~\ref{fig:KS-diagram}
shows four branches representing $r=r_{\text{s}}$ in total. For~$\tilde{k}=0$
(i.e., $\zeta=1$), the~hyperbolic branches \eqref{eq:hyperbolae}
degenerate into two straight lines, $T=\pm X$, as~is expected for
the Schwarzschild metric. In~the limit of $\tilde{k}\rightarrow0$,
Region (VI), which sandwiches within the four hyperbolic branches,
shrinks and disappears.
\item Region (I) refers to $r>r_{\text{s}}$ (the ``exterior''); Region
(II) refers to $0<r<r_{\text{s}}$. 
\item Regions (III) and (IV) are double copies of Regions (I) and (II) respectively,
by flipping the sign of the KS coordinates, viz. $(T,X)\leftrightarrow(-T,-X)$.
Regions (Va) and (Vb) are unphysical, viz. $r<0$. 
\item Region (VI) generally contains curvature singularities, with~the Kretschmann
scalar generally diverging on the hyperbolic branches given
in \eqref{eq:hyperbolae}. The~``gulf'' represented by Region (VI)
is a new feature of the asymptotically flat Buchdahl-inspired spacetimes. {The
 Kretschmann invariant is reproduced in Appendix \ref{sec:Kretschmann-scalar}.}
\end{enumerate}

\subsection*{Areal Radius}

The $\zeta$--KS diagram, depicted in Figure~\ref{fig:KS-diagram},
is the maximal analytic extension of the \emph{special} Buchdahl-inspired
metric, given in Equation~\eqref{eq:B-metric-new}. Similar to the usual
KS diagram for the Schwarzschild metric, our $\zeta$--KS diagram reveals
a double-cover, comprising Regions (III) and (IV).

In~\citep{2023-WH}, one of the present authors showed that, in~certain~situations,
Region (I) and its double-cover Region (III) can be further split.
When this splitting occurs, the~physical singularities on the hyperbolic
branches of $r=r_{\text{s}}$ are shielded from an observer situated
at spatial infinity, and~the double covers that are connected to spatial
infinity can be seamlessly ``glued'' together to form a two-way
traversable wormhole. This procedure was carried out in~\citep{2023-WH}.
However, we shall briefly review the analysis here, for~self-consistency
and~self-completeness.

\begin{figure}[H]
\includegraphics[scale=0.78]{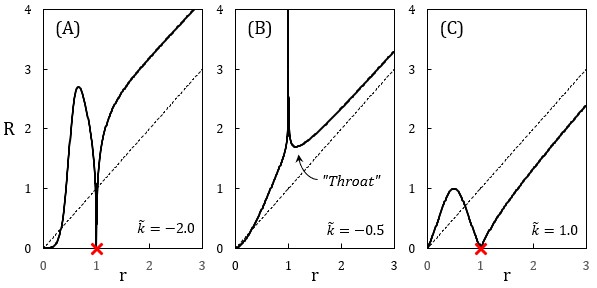}\vspace{-3pt}
\caption{\label{fig:R(r)-special}$R$ vs. $r$ for the \emph{special} Buchdahl-inspired
metric; using $r_{\text{s}}=1$. Panel (\textbf{B}), representative of $\tilde{k}\in(-1,0)$,
yields a minimum for $R(r)$ and corresponds to a wormhole. Panels
(\textbf{A},\textbf{C}), representative of $\tilde{k}\in(-\infty,-1)\cup(0,+\infty)$,
show a monotonic behavior for $R(r)$ in the \textquotedblleft exterior\textquotedblright ,
$r>r_{\text{s}}$.}
\end{figure}

From Equation~\eqref{eq:B-metric-new}, the~areal radius is given by\vspace{-6pt}
\begin{equation}
R(r)=r\left|1-\frac{r_{\text{s}}}{r}\right|^{\frac{1}{2}\left(1+\frac{\tilde{k}-1}{\zeta}\right)}
\end{equation}
Figure~\ref{fig:R(r)-special} depicts $R$ as function for $r$ for
various values of $\tilde{k}$. Furthermore, since\vspace{-6pt}
\begin{equation}
\frac{dR}{dr}=\frac{r-\left(\frac{1}{2}-\frac{\tilde{k}-1}{2\zeta}\right)r_{\text{s}}}{r-r_{\text{s}}}\left|1-\frac{r_{\text{s}}}{r}\right|^{\frac{1}{2}\left(1+\frac{\tilde{k}-1}{\zeta}\right)},
\end{equation}
the equation $dR/dr=0$ has a single root, given by\vspace{-6pt}
\begin{equation}
r_{*}=\frac{r_{\text{s}}}{2}\,\left(1-\frac{\tilde{k}-1}{\zeta}\right)\,.\label{eq:r-star}
\end{equation}
This root lies in the range of $(r_{\text{s}},\infty)$ as a local
minimum if $\tilde{k}\in(-1,0)$ and in the range of $(0,r_{\text{s}})$
as a local maximum if $\tilde{k}\in(-\infty,-1)\cup(0,+\infty)$.
We refer the reader to Figure~\ref{fig:R(r)-special} for more details,
and below, we briefly analyze both~cases.
\vspace{-3pt}

\section{Naked Singularity: \boldmath{$\ $Case $\tilde{k}\in(-\infty,-1)\cup(0,+\infty)$}}

\label{sec:naked}

This case corresponds to Figure~\ref{fig:R(r)-special}A,C. The ``exterior'' Region (I) forms one continuous sheet, and its double cover Region (III) forms another continuous~sheet.

\begin{figure}[H]
\includegraphics[scale=0.65]{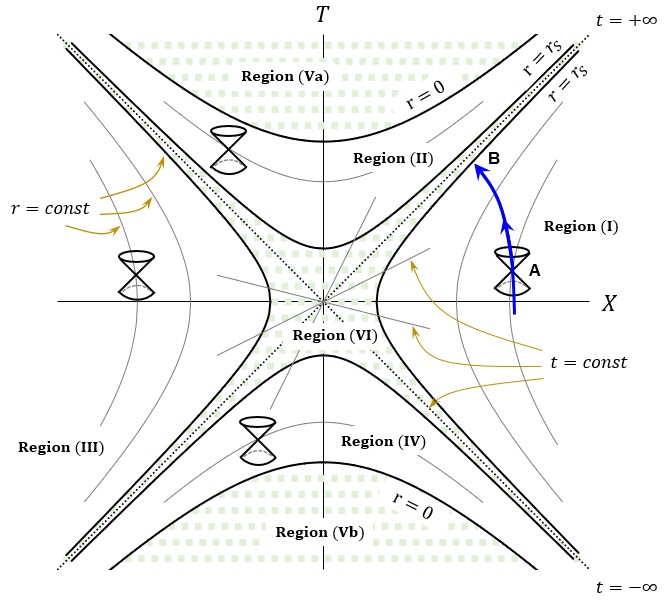}\vspace{-3pt}
\caption{\label{fig:KS-diagram-NS}The case of naked singularities, $\tilde{k}\in(-\infty,-1)\cup(0,+\infty)$.
On the infalling radial timelike trajectory (blue line), a~particle
in Region (I) eventually hits the naked singularity at $r=r_{\text{s}}$.}
\end{figure}

We plot a timelike trajectory $A\rightarrow B$ in Figure~\ref{fig:KS-diagram-NS}.
In Region (I), an~infalling traveller reaches the
singularity at point B after a finite amount of proper time. Note
that in the view of an observer who stays at rest at spatial infinity
in Region (I), it takes a finite amount of time $t$ for the traveller
to hit the singularity. This is \emph{in contrast to} the Schwarzschild
metric, where an observer from afar will never witness the traveller
reaching the horizon at $r=r_{\text{s}}$.

What happens to the traveller after reaching the singularity remains
an open question with several possibilities. There are at least three
scenarios to consider: (i) the traveller might come to a halt at $r=r_{\text{s}}$
without further motion; (ii) the traveller might enter the ``gulf''
region, viz. Region (VI); or (iii) the traveller might directly pass
into Region (II) and continue heading towards the singularity at $r=0$.
Regardless of the outcome, we do not concern ourselves with this case. 
Our focus is directed towards the wormhole scenario, which
will be presented~next.
\vspace{-3pt}

\section{Wormhole: \boldmath{$\ $Case $\tilde{k}\in(-1,0)$}}\label{sec:CTC}

\begin{figure}[H]
\includegraphics[scale=0.45]{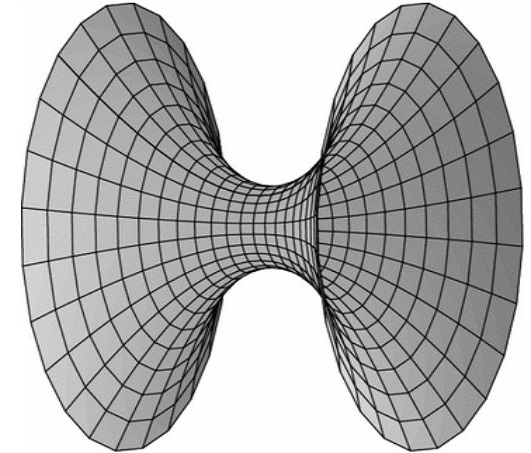}\vspace{0.5cm}\vspace{-6pt}
 \caption{\label{fig:Embed}Embedding diagram of a typical traversable wormhole. The wormhole \textquotedblleft throat\textquotedblright{} is depicted
horizontally to be compatible with Figure~\ref{fig:KS-diagram-WH} on
page 9, with~the right mouth corresponding
to Region (Ia) and the left mouth Region (IIIa). Note, however, that
the embedding diagram is a \textquotedblleft snapshot\textquotedblright{}
at a fixed timeslice $T$ (while the azimuth direction is made explicit),
whereas the $\zeta$--KS shows the full \textquotedblleft evolution\textquotedblright{}
in the $T-$direction (with both the polar and azimuth angles being suppressed).}
\end{figure}

Panel (B) in Figure~\ref{fig:R(r)-special} is representative of this
case. The~areal radius $R$ exhibits a minimum at $r_{*}=\frac{r_{\text{s}}}{2}\,\left(1-\frac{\tilde{k}-1}{\zeta}\right)$
in the ``exterior'' region, viz. $r_{*}>r_{\text{s}}$, per Equation~\eqref{eq:r-star}. In~Ref.~\citep{2023-WH}, this fact was employed to construct a Morris--Thorne--Buchdahl (MTB) wormhole with its throat located at $r_{*}$. A~schematic depiction of the wormhole at a specific time-slice (at a given T), with~the azimuth angle $\varphi$ shown and the polar angle $\theta$ suppressed, is shown in Figure~\ref{fig:Embed}. We refer the reader to Ref.~\citep{2023-WH} for~details.

\begin{figure}[H]
\includegraphics[scale=0.68]{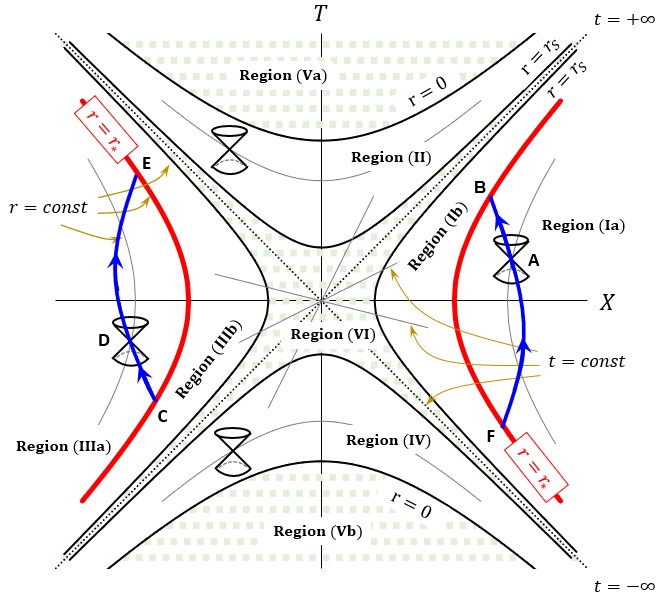}\vspace{-6pt}
\caption{\label{fig:KS-diagram-WH}The case of the traversable wormhole, with $\tilde{k}\in(-1,0)$. The~wormhole throat is depicted by the red lines, which further split Region (I) into (Ia) and (Ib), and~Region
(III) into (IIIa) and (IIIb). On~the radial trajectory $A\rightarrow B\equiv C\rightarrow D\rightarrow E\equiv F\rightarrow A$,
an infalling traveller in sub-region (Ia) first enters the wormhole
mouth at point B then traverses into sub-region (IIIa) by emerging
at the other mouth at point C (hence, on~outgoing motion). As~the
two red lines are \textquotedblleft glued\textquotedblright{} together
to form a wormhole that connects sub-region (Ia) and sub-region (IIIa),
the pair of antipodal points B and C represent a single spacetime event.
Likewise, the~pair of antipodal points E and F correspond to a single
spacetime event. The~segment $C\rightarrow D\rightarrow E$ progresses
\emph{backward} in time as compared with the segment $F\rightarrow A\rightarrow B$.}
\end{figure}

The $\zeta$--KS diagram exhibits an additional feature: in Figure~\ref{fig:KS-diagram-WH}, the~loci where $r=r_{*}$, representing the local minimum areal radius,
are depicted as two thick red hyperbolic branches. These branches
partition the ``exterior'' Region (I) into two sub-regions, denoted
as (Ia) and (Ib), while the double-cover Region (III) is also divided
into two sub-regions, (IIIa) and (IIIb). The~two asymptotically flat
sheets in sub-region (Ia) and sub-region (IIIa) are seamlessly connected
or ``glued'' together along the two thick red hyperbolic branches
(as well as along the polar angle $\theta$ and the azimuth angle
$\varphi$ of the two-sphere) to form a four-dimensional~wormhole.

\vspace{-10pt}

It is essential to note that, according to Equations~\eqref{eq:T-X-1}
and \eqref{eq:T-X-2}, each spacetime event $(t,r)$ with $r$ in
the range of $(r_{\text{s}},\infty)$ corresponds to two antipodal
points $(T,X)$ and $(-T,-X)$ of the KS coordinates. This duplication
(or double degeneracy) is nothing but a double copy of exterior sheets.
However, in~general, these two points correspond to distinct spacetime
events which occur in two separate sheets. \emph{Only along the loci
$r=r_{*}$} do the sheets become ``glued'' together, forming a wormhole
throat with the two thick red hyperbolic branches as the two mouths.
\emph{Along the throat, the~two antipodal points correspond to the
same spacetime event.} In other words, in~Figure~\ref{fig:KS-diagram-WH},
point B and point C are identical, as~are points E and F. Note
that this identification does \emph{not} apply, for~instance, to~the
pair of antipodal points A and D, as~\emph{these points stay off the
throat}. That is to say,  despite
sharing the same value of $r$ and the same value of $t$, point A
and point D represent two independent events that take place in two
separate spacetime~sheets.

\begin{figure}[H]
\includegraphics{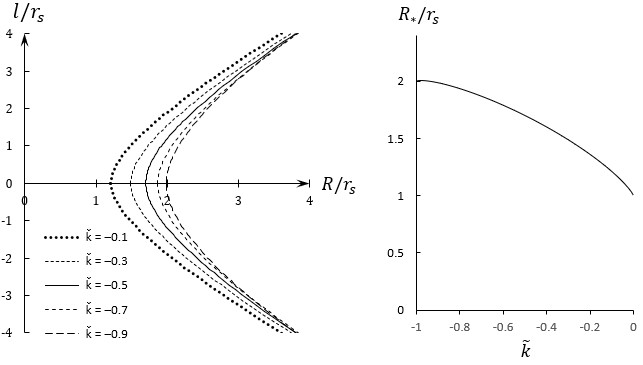}
 \caption{{\label{fig:Proper-radial-coordinate}(\textbf{Left}): Proper radial coordinate as a function of areal radius for various
value of $\tilde{k}\in(-1,0)$, per Equations~\eqref{eq:proper-l} and \eqref{eq:areal-radius}.
Each curve is} {\emph{vertical}}{{}
at $l=0$, the~location of a wormhole throat. (\textbf{Right}): The size
of the throat as a function of $\tilde{k}\in(-1,0)$.}}
\end{figure}

{The identification of antipodal points} {\emph{on
the throat}} {can be illuminated by examining the
proper radial coordinate. This coordinate is expressible using Gaussian
hypergeometric functions, with~$\zeta:=\sqrt{1+3\tilde k^2}$ and $B:=(\tilde k -1)/\zeta$, as~derived in~\citep{2023-WH}:
\begin{align}
l(R) & =\pm\int_{R_{*}}^{R}\frac{dR}{\sqrt{1-\frac{b(R)}{R}}}\\
 & =\pm\frac{\zeta\,r_{\text{s}}}{1+\frac{B}{2}}\times\biggl[y^{1+\frac{B}{2}}\,_{2}F_{1}\Bigl(2,1+\frac{B}{2};2+\frac{B}{2};y\Bigr)-y_{*}^{1+\frac{B}{2}}\,_{2}F_{1}\Bigl(2,1+\frac{B}{2};2+\frac{B}{2};y_{*}\Bigr)\biggr]\label{eq:proper-l}
\end{align}
where the areal radius is given by
\begin{equation}
R(y)=\zeta\,r_{\text{s}}\frac{y^{\frac{1}{2}(B+1)}}{1-y}\label{eq:areal-radius}
\end{equation}
Figure~\ref{fig:Proper-radial-coordinate} illustrates the proper radial coordinate for various values of $\tilde{k}$ within the range $(-1,0)$. For~instance, when $\tilde{k}=-0.5$, it corresponds to
$B\approx-1.134$, $y_{*}\approx0.0628$, $r_{*}\approx1.14\,r_{\text{s}}$,
$R_{*}\approx1.7\,r_{\text{s}}$.}

{In Figure~\ref{fig:Proper-radial-coordinate}, for~
a given $\tilde{k}\in(-1,0)$, the~curve of $l(R)$ consists of two
semi-infinite segments: The segment with $l>0$ corresponds to Region
(Ia), whereas the segment with $l<0$ to Region (IIIa) in the $\zeta$--KS
diagram of Figure~\ref{fig:KS-diagram-WH}. These two segments are connected
at the throat, marked by $l=0$ and $R=R_{*}$ (the minimum in the
real radius), at~which point the curve $l(R)$ is} {\emph{vertical}}{.
It is only at this specific juncture that the two asymptotic flat sheets,
viz. Regions (Ia) and (IIIa), ``touch'' each other, allowing a transition
from one sheet to another and enabling the identification of antipodal
points to take effect.}

{The coordinate $l$ covers the entire range $(-\infty,+\infty)$.
A geodesic running across the throat is thus} {\emph{complete}}{,
without encountering any physical singularities. As~the MTB wormhole
consists of Regions (Ia) and (IIIa), the~physical singularities located
at $r=r_{\text{s}}$ are insulated within Regions (Ib) and (IIIb),
which are not components of the wormhole.}

Now, consider an intrepid traveller tracking an infall starting from
point~A:
\begin{itemize}
\item The traveller enters the right mouth of the wormhole at point B after
a finite amount of proper time. From~the perspective of an observer
at rest in Region (Ia), it also takes a \emph{finite} amount of time
for the traveller to reach point B.
\item Subsequently, the~traveller emerges from the left mouth of the wormhole
at point C (note: point B and point C are identical!). They then ascend
the potential well (viz. \emph{increasing} $r$), moving toward point
D. It is important to note that, relative to the observer in Region
(Ia), the~traveller appears to move \emph{backward} in time, as~the
$t$ coordinate \emph{decreases} from point C to point D.
\item If the traveller chooses to fall back into the left mouth, they will
re-enter it at point E. They will then re-emerge at the right mouth
at point F (which is identical to point E!). Notably, upon~re-emergence,
they are once again moving \emph{forward} in time.
\item At this stage, the~traveller can choose to proceed to point A, thereby
\emph{completing a closed timelike loop}.
\end{itemize}

The traveller's ability to complete such a loop relies on the \emph{two
reversals of time direction}, occurring each time they enter a wormhole
mouth and emerges from the other~mouth.

\section{Discussions and~Summary}\label{sec:discussions}

In the preceding section, we have constructed a closed timelike curve
(CTC) by having a traveller pass through the wormhole throat twice
in succession. During~each passage, the~traveller experiences a reversal
in the time direction \emph{with respect to} an observer at rest.
This construction is succinctly captured in Figure~\ref{fig:KS-diagram-WH},
where the closed path $A\rightarrow B\equiv C\rightarrow D\rightarrow E\equiv F\rightarrow A$
forms a~CTC.
\begin{enumerate}
\item It is worth highlighting that the CTC we have described does not require
having one wormhole mouth move at high speed or be located near a supermassive
object to accumulate time dilation, as~popularized in~\citep{MorrisThorne-1988-2,Visser-book,Lobo-2008}.
To the best of our knowledge, the~CTC presented herewith has not been
documented in the existing literature.
\item Our association of a pair of antipodal points (such as B and C) with
a single spacetime event was inspired by a similar construction proposed
by Pop\l{}awski. In~\citep{Poplawski-2010}, he revisited the Einstein--Rosen
(ER) bridge and identified two antipodal points \emph{on the horizon}
with a single spacetime event. The~ER bridge was interpreted as a
``Schwarzschild wormhole'' connecting the two exterior sheets joined
\emph{at the horizon}. However, the~ER bridge encounters various issues
related to traversability, stability, and~a problematic thin-shell
mass distribution at the horizon. (It is also worth noting that Pop\l{}awski
did not suggest the possibility of CTCs in his work.)
\item The Morris--Thorne--Buchdahl wormhole, developed in~\citep{2023-WH}
and briefly summarized in this paper, appears to avoid the issues
faced in Pop\l{}awski's work. Setting aside the \emph{physical} questions
regarding causality violations and time-travel-related paradoxes,
our construction of a CTC appears to be \emph{mathematically} consistent.
\item Central to our approach is the identification of any given pair of
antipodal points on the loci of minimum areal radius, denoted as $r=r_{*}$,
with a single spacetime event. This identification becomes evident
through Equations~\eqref{eq:T-X-1} and \eqref{eq:T-X-2}, as~
well as in the KS diagrams employed for the Schwarzschild metric (used
in Pop\l{}awski's work) and our \emph{special} Buchdahl-inspired
metric in Figure~\ref{fig:KS-diagram-WH}.
\item Embedding diagrams, such as the one in Figure~\ref{fig:Embed},
depict a \textquotedblleft snapshot\textquotedblright{} of the wormhole
at a fixed timeslice, without~illustrating the evolution of timelike
trajectories. As~such, the~use of embedding diagrams may obscure
the identification of spacetime events on the throat, a~procedure
we carried out in this paper. In~this regard, the~decisive advantage
of KS diagrams is their ability to reveal the full causal structure
of spacetime, making this identification transparent.
\item \textls[-15]{Let us draw a comparison between our CTC and the Deutsch-Politzer (DP) time~machine.}
\begin{itemize}
\item In~\citep{Deutsch-1991,Politzer-1992}, a~DP space is created from
a two-dimensional Minkowski spacetime by making two finite-size ``space-like''
cuts and gluing the edges of the cuts, effectively forming a ``handle''
which connects two space-like regions and creates a time machine.
The essence of a DP time machine is that \emph{the spacetime topology
is altered} {[}Note that the DP space has singularities; to exorcise
them, in~\citep{Krasnikov-1998,Krasnikov-1995} Krasnikov performed
a conformal transformation to send the singular points away to infinity{]}.
\item Our CTC construction shares both analogies and differences with the
DP time machine. The~$\zeta$--KS diagram of the Buchdahl-inspired
vacuum (Figure \ref{fig:KS-diagram-WH}) is a two-dimensional Minkowski
spacetime (modulo a Weyl transformation). For~$\tilde{k}\in(-1,0)$,
the hyperbolic branches of $r=r_{*}$ are glued to form a portal between
the two time-reversed sheets, viz. Regions (Ia) and (IIIa). \emph{In
this regard, akin to the DP space, our construction is a concrete
realization of a time machine induced by an alteration in the topology
of spacetime.}
\item Unlike the deliberate surgery employed in the DP space, the~alteration
of the topology in the Buchdahl-inspired vacuum occurs naturally,
driven by the fourth-derivative dynamics of pure $\mathcal{R}^{2}$
gravity. {(}Also, similar to Krasnikov's work~\citep{Krasnikov-1998,Krasnikov-1995},
Regions (Ia) and (IIIa) in our $\zeta$--KS diagram are devoid of (physical)
singularities.{)} A comprehensive discussion of their commonalities
and differences exceeds the scope of this paper.
\end{itemize}
\item Our CTC construction is not confined solely to the MTB wormholes of
pure $\mathcal{R}^{2}$ gravity. It is also applicable to two-way
traversable wormholes in Brans--Dicke (BD) gravity, which share a similar
Kruskal--Szekeres diagram, as~demonstrated in Ref.~\citep{2023-WEC}
by one of the~authors.
\begin{itemize}
\item {More generally, the~Brans wormhole, first established
by Agnese and La Camera for BD gravity in~\citep{Agnese-1995,Agnese-2001},
encompasses the MTB wormhole in $\mathcal{R}^{2}$ gravity~\citep{2023-WEC}.
It is known that a static vacuum solution of any $f(R)$ gravity cannot
host a twice asymptotically flat wormhole~\citep{Bronnikov-2010}.
However, a~cut-and-paste procedure can be employed to generate such
a wormhole, a~technique that underlies the creation of the Brans wormhole
\citep{Agnese-1995,Agnese-2001,Nandi-1997,Faraoni-2012}.}
\item {The MTB wormhole satisfies the four ``traversability-in-principle''
criteria laid out by Morris and Thorne~\citep{MorrisThorne-1988-1}.
To be considered ``usable'', the~tidal forces should remain finite.
We have computed the tidal forces in Appendix \ref{sec:Tidal-forces}.
Our findings indicate that despite jumps in higher derivatives across
the throat in the metric components, the~tidal forces remain finite
throughout the two asymptotically flat spacetime sheets.}
\end{itemize}
\end{enumerate}

\textls[-25]{In conclusion, the~presence of CTCs is an extremely subtle issue which
needs to be handled with great caution, due to the association with
time travel paradoxes, such as the classical consistency paradoxes
and causal loops~\citep{Lobo-2008}. Much has been written on the
resolution to the paradoxes associated with CTCs, such as the Principle
of Self-Consistency~\citep{Earman,Echeverria,NovikovCTC} and the
Chronology protection conjecture~\citep{hawking} (we refer the reader
to~\citep{Lobo-2008} for more details). The~issue of CTCs is an extremely
fascinating research topic and is essentially useful as ``gedanken-experiments''
that force us to confront the foundations of general relativity, and~
its modifications, and~extract clarifying views.}



\vspace{6pt} 
\authorcontributions{All the authors have substantially contributed to the present work. All~authors have read and agreed to the published version of the~manuscript.}

\funding{This research was funded by the Funda\c{c}\~{a}o para a Ci\^{e}ncia e a Tecnologia (FCT) from the research grants UIDB/04434/2020, UIDP/04434/2020 and CERN/FIS-PAR/0037/2019 and PTDC/FIS-AST/0054/2021.}

\institutionalreview{Not applicable.}

\informedconsent{Not applicable.}

\dataavailability{Not applicable.} 



\acknowledgments{The authors thank the anonymous referees for insightful and helpful comments. H.K.N. wishes to thank Tiberiu Harko, Mustapha Azreg-A\"{i}nou, and~Nicholas Buchdahl. F.S.N.L. acknowledges support from the Funda\c{c}\~{a}o para a Ci\^{e}ncia e a Tecnologia (FCT) Scientific Employment Stimulus contract with reference CEECINST/00032/2018, and~funding from the research grants UIDB/04434/2020, UIDP/04434/2020 and CERN/FIS-PAR/0037/2019.}

\conflictsofinterest{The authors declare no conflict of~interest.} 
\appendixtitles{yes} 
\appendixstart
\appendix
\section[\appendixname~\thesection]{{\label{sec:Kretschmann-scalar}\ Kretschmann~Scalar}}

{The Kretschmann invariant has been computed in~\citep{Bronnikov-1997,2023-WEC}
\begin{align}
K & :=\mathcal{R}_{\mu\nu\rho\sigma}\mathcal{R}^{\mu\nu\rho\sigma}\\
 & =\left|1-\frac{r_{\text{s}}}{r}\right|^{-\frac{2}{\zeta}(\tilde{k}-1+2\zeta)}\ \frac{r_{\text{s}}^{2}}{r^{6}}\,\biggl(6\mathfrak{A}-2\mathfrak{B}\,\frac{r_{\text{s}}}{r}+\frac{\mathfrak{C}}{4}\,\frac{r_{\text{s}}^{2}}{r^{2}}\biggr)
\end{align}
in which $A:=\frac{\tilde{k}+1}{\zeta}$, $B:=\frac{\tilde{k}-1}{\zeta}$,
$\zeta:=\sqrt{1+3\tilde{k}^{2}}$, whereas
\begin{align}
\mathfrak{A} & =A^{2}+B^{2}\\
\mathfrak{B} & =A^{2}(A-2B+3)-B(B-1)(B-2)\\
\mathfrak{C} & =(A+1)A^{2}(A-2B+3)+(3A^{2}+B^{2}-2B+3)(B-1)^{2}
\end{align}
}
\begin{itemize}
\item {At $\tilde{k}=0$: $\zeta=1$, $\mathfrak{A}=2$,
$\mathfrak{B}=12$, $\mathfrak{C}=48$, giving $K=12\frac{r_{\text{s}}^{2}}{r^{6}}$
in agreement with the standard result fort the Schwarzschild metric.}
\item {At $\tilde{k}=-1$: $\zeta=2$, $\mathfrak{A}=1$,
$\mathfrak{B}=6$, $\mathfrak{C}=24$, giving $K=\ 6\frac{r_{\text{s}}^{2}}{r^{6}}$.}
\item {For $\tilde{k}\neq0$ and $\tilde{k}\neq-1$, as~
$\tilde{k}-1+2\zeta>0$, $K$ generally diverges at $r=r_{\text{s}}$
and $r=0$.\vskip8pt}
\end{itemize}

\section[\appendixname~\thesection]{{\label{sec:Tidal-forces}\ Tidal Forces in MTB and
Brans Wormholes}}

{In addition to the four ``traversability-in-principle''
criteria~\citep{MorrisThorne-1988-1}, it is desirable to require
that the metric components be at least twice-differentiable in terms
of the radial coordinate~\citep{Visser-1997}. This ``usability''
requirement} {\emph{may}} { be justified
since the tidal forces, which are the physical quantities of concern,
involve the first and second derivatives of the redshift and shape
functions (as we shall see momentarily). Let us compute the tidal
forces for the Morris--Thorne--Buchdahl (MTB) wormholes and the Brans
wormholes.\vskip4pt}

{We shall adopt Morris--Thorne (MT)'s exposition~\citep{MorrisThorne-1988-1}.
For the MT ansatz
\begin{equation}
ds^{2}=-e^{2\Phi(R)}dt^{2}+\frac{dR^{2}}{1-\frac{b(R)}{R}}+R^{2}d\Omega^{2}\label{eq:MT-ansatz}
\end{equation}
the radial and lateral tidal forces are proportional to (with primes
denoting derivatives with respect to $R$):\vspace{-6pt}
\begin{align}
\left|\mathcal{R}_{\hat{1}'\hat{0}'\hat{1}'\hat{0}'}\right| & =\left|\left(1-\frac{b}{R}\right)\left(-\Phi''+\frac{b'R-b}{2R(R-b)}\Phi'-\Phi'^{2}\right)\right|\\
 & =\left|\left(1-\frac{b}{R}\right)\left(-\Phi''-\Phi'^{2}\right)-\left(1-\frac{b}{R}\right)'\frac{\Phi'}{2}\right|\label{eq:tidal-1}\\
\left|\mathcal{R}_{\hat{2}'\hat{0}'\hat{2}'\hat{0}'}\right| & =\left|\frac{\gamma^{2}}{2R^{2}}\left[\frac{v^{2}}{c^{2}}\left(b'-\frac{b}{R}\right)+2(R-b)\phi'\right]\right|\\
 & =\left|\frac{\gamma^{2}}{2R}\left[-\frac{v^{2}}{c^{2}}\left(1-\frac{b}{R}\right)'+2\left(1-\frac{b}{R}\right)\phi'\right]\right|\label{eq:tidal-2}
\end{align}
These expressions are Equations~(49) and (50) 
 in the MT paper~\citep{MorrisThorne-1988-1}.
It thus appears desirable to impose twice-differentiability on $\Phi$
and first-differentiability on $b$ as functions of $R$.\vskip4pt}

{Let us apply these formulae for the asymptotically
flat Buchdahl-inspired solution~\citep{2023-WH} and the Campanelli--Lousto
solution~\citep{2023-WEC,Agnese-1995}. Both of these solutions can
be cast in the MT ansatz, Equation~\eqref{eq:MT-ansatz}, in~which the
redshift and shape functions are~\citep{2023-WH}
\begin{align}
e^{2\Phi(R)} & =y^{A}\\
1-\frac{b(R)}{R} & =\frac{(1-B)^{2}}{4y}\left(y-\frac{B+1}{B-1}\right)^{2}
\end{align}
in which the auxiliary variable $y:=\left(1-\frac{r_{\text{s}}}{r}\right)^{\zeta}\in(0,1)$
and the areal radius $R$ is expressed~as
\begin{equation}
R(y)=\zeta r_{\text{s}}\frac{y^{\frac{1}{2}(B+1)}}{1-y}\,.
\end{equation}
For the Campanelli--Lousto solution, $A$ and $B$ are two independent
parameters, while $\zeta:=1$~\citep{2023-WEC,Agnese-1995}. For~the
asymptotically flat Buchdahl-inspired solution, $A$ and $B$ are
related by the following definitions~\citep{2023-WH}
\begin{equation}
A:=\frac{\tilde{k}+1}{\zeta};\ \ B:=\frac{\tilde{k}-1}{\zeta};\ \ \zeta:=\sqrt{1+3\tilde{k}^{2}}
\end{equation}
As demonstrated in Ref.~\citep{2023-WH}, for~$\tilde{k}\in(-1,0)$,
the areal radius has a minimum at
\begin{equation}
y_{*}:=\frac{B+1}{B-1}\in(0,1)\label{eq:y_star}
\end{equation}
This is the location of the throat, and~the variable $y$ for} {\emph{both}}{{}
of the two spacetime sheets are in the range $[y_{*},1)$.\vskip4pt}

{Direct calculations yield:
\begin{equation}
\left(1-\frac{b}{R}\right)'=\frac{(1-B)y^{-\frac{1}{2}(B+3)}(1-y)^{2}}{2\zeta r_{\text{s}}}\left(y+\frac{B+1}{B-1}\right)
\end{equation}
\begin{equation}
\left(1-\frac{b}{R}\right)\Phi'=\frac{A(1-B)y^{-\frac{1}{2}(B+3)}(1-y)^{2}}{4\zeta r_{\text{s}}}\left(y-\frac{B+1}{B-1}\right)
\end{equation}
\begin{align}
\left(1-\frac{b}{R}\right)\left(-\Phi''-\Phi'^{2}\right)-\left(1-\frac{b}{R}\right)'\frac{\Phi'}{2} & =\frac{A(2+A-B)y^{-B-2}(1-y)^{3}}{4\zeta^{2}r_{\text{s}}^{2}}\times\nonumber \\
 & \ \ \ \ \ \ \ \ \ \ \ \ \ \ \ \left(y-\frac{B-A+2}{B-A-2}\right)\label{eq:tmp}
\end{align}
Since all these expressions do not contain any singularity for $y\in[y_{*},1)$,
}{\emph{the tidal forces in Equations~\eqref{eq:tidal-1}
and \eqref{eq:tidal-2} are finite everywhere in this range}}{.
It is important to note that the tidal forces may have opposite signs
in the two sheets across the throat, but~}{\emph{they
remain finite}}{{} on the two asymptotically flat sheets
that form the MTB wormhole.}

{Furthermore, it should be noted that all the calculations
between Equations~\eqref{eq:y_star} and~\eqref{eq:tmp} are extendible
to the Campanelli--Lousto solution provided that $B<-1$, in~which
case a Brans wormhole has been established to exist~\citep{2023-WEC,Agnese-1995}.
Our conclusion regarding the finite tidal forces is therefore applicable
to Brans wormholes as well.}


\begin{adjustwidth}{-\extralength}{0cm}

\reftitle{References}

\PublishersNote{}
\end{adjustwidth}
\end{document}